\documentclass[secnumarabi,amssymb,nobibnotes,aps,prc]{revtex4}

\usepackage{epsfig}
\usepackage{graphicx}
\usepackage{amsmath}
\usepackage{hyperref}

\begin{document} 

\title{Incoherent $\rho$ meson photoproduction in ultraperipheral nuclear collisions at the CERN Large Hadron Collider}

\author{V. Guzey}

\author{E. Kryshen}

\author{M. Zhalov}
\affiliation{National Research Center ``Kurchatov Institute'', Petersburg Nuclear Physics Institute (PNPI), Gatchina, 188300, Russia}


\begin{abstract} 

Using the Gribov--Glauber model for photon--nucleus scattering and a generalization 
of the vector meson dominance model for the hadronic structure of the photon,  
we make predictions for the cross section of incoherent 
$\rho$ photoproduction in Pb-Pb ultraperipheral collisions (UPCs) in the Large Hadron Collider kinematics.
We find that the effect of the inelastic nuclear shadowing is significant 
and leads to an additional 25\% suppression of the incoherent cross section.
Comparing our predictions to those of the STARlight Monte Carlo framework, 
we observe very significant differences.

\end{abstract}

\maketitle 

\section{Introduction}
\label{sec:intro}

Ultraperipheral collisions (UPCs) of heavy ions at the Relativistic Heavy Ion Collider (RHIC) 
and the Large Hadron Collider (LHC) give an opportunity to explore high-energy nuclear physics 
with beams of quasi-real photons~\cite{Baltz:2007kq}. Indeed, 
UPCs of relativistic ions are 
characterized by large transverse distances $b$ (impact parameters) 
between the centers of colliding nuclei of radii $R_A$ and $R_B$, $b \gg R_A+R_B$, so that
the strong nucleus--nucleus interaction is suppressed leading to dominance of the 
photon--nucleus interactions involving quasi-real photons emitted by the colliding nuclei. 
These photons have a wide energy spectrum extending into a TeV-range for the LHC beam 
energies (in the target nucleus rest frame) and the high flux scaling as $Z^2$ ($Ze$ is the 
electric charge of the photon-emitting nucleus).
Among many exciting directions of UPC studies, 
coherent and incoherent photoproduction of light vector mesons 
on nuclei allow one to investigate soft meson--nucleus interactions at high energies in the 
kinematic domain unavailable with fixed nuclear targets.
While it is generally understood and accepted that the strong suppression of cross 
sections of high-energy soft 
hadron--nucleus scattering is due to the nuclear shadowing effect arising as the result 
of multiple interactions of the projectile with target 
nucleons~\cite{Glauber:1955qq,Glauber:1970jm,Gribov:1968jf}, the practical implementation of 
this mechanism in coherent and incoherent photoproduction of light vector mesons in UPCs 
differs in the literature~\cite{Frankfurt:2002wc,Frankfurt:2002sv,Rebyakova:2011vf,Frankfurt:2015cwa,Guzey:2016piu,
Goncalves:2011vf,Santos:2014vwa,Goncalves:2017wgg,Klein:1999qj,Klein:2016yzr,Cepila:2018zky,Khoze:2019xke}. 
In particular, the open questions include the generalization of the naive vector meson 
dominance (VMD) model to the case of photon--nucleus interactions, the comparatively large 
uncertainty in the experimentally determined magnitude of the vector meson--nucleon cross section,
and the size and energy-dependence of the inelastic 
(Gribov) shadowing correction. Also, attempts to calculate the nuclear cross sections with a precision
better than $10\%$ immediately raise the issues of the dependence of the nuclear cross section 
on the choice of the nuclear wave function, the role of short-range nucleon--nucleon 
correlations, and the finite range (radius) of hadron--nucleon interactions.

This paper extends our earlier 
work~\cite{Frankfurt:2015cwa} on coherent $\rho$ meson photoproduction 
on nuclei in UPCs of heavy ions in the kinematics of the LHC
to the case of incoherent $\rho$ photoproduction 
(the target nucleus breaks up), where differences among scarce theoretical predictions 
seem to be very significant.
Also, such an analysis is topical in view of anticipated LHC experimental data on this process.
In particular, we calculate the incoherent $Pb Pb \to \rho Pb A^{\prime}$ UPC cross 
section ($A^{\prime}$ stands for the products of dissociation of the nuclear target) as a 
function of the rapidity $y$ at the invariant center-of-mass energy $\sqrt{s_{NN}}=5.02$ TeV
and as a function of $\sqrt{s_{NN}}$ at the central rapidity $y=0$. We also present our results 
for the incoherent photoproduction cross section $\gamma Pb \to \rho A^{\prime}$ as a 
function of the photon--nucleon energy $W_{\gamma p}$. 

Our results show that within the adopted model for the hadronic structure of the incident photon, 
the effect of the inelastic nuclear shadowing in the incoherent cross sections is significant. This 
observation complements a similar conclusion in the coherent case~\cite{Frankfurt:2015cwa}, 
which was confirmed by a comparison of our calculations with the ALICE measurements~\cite{Adam:2015gsa,Acharya:2020sbc}.
  We also find very significant differences of our results for the incoherent photoproduction cross 
section from predictions of the STARlight Monte Carlo framework, which is 
frequently used in the UPC experimental data processing and analysis~\cite{Klein:2016yzr}.

The remainder of the paper is structured as follows.
In Sec.~\ref{sec:expressions}, we present expressions for the UPC, incoherent and coherent 
nuclear photoproduction cross sections, which we used in our analysis. 
Our predictions for the resulting cross sections in the LHC kinematics, their discussion and comparison to
the STARlight model are given
in Sec.~\ref{sec:results}. Finally, we summarize and draw conclusions in Sec.~\ref{sec:summary}.

\section{Incoherent vector meson photoproduction in heavy-ion ultraperipheral collisions}
\label{sec:expressions}

In the case of incoherent photoproduction of $\rho$ mesons in 
symmetric UPCs of ions $A$ and using the 
equivalent photon approximation~\cite{Budnev:1974de},
the UPC cross section can be written in the following form~\cite{Baltz:2007kq}
\begin{equation}
\frac{d \sigma_{AA \to \rho AA^{\prime}}}{dy}=N_{\gamma/A}(y) \sigma_{\gamma A \to \rho A^{\prime}}(y)
+N_{\gamma/A}(-y) \sigma_{\gamma A \to \rho A^{\prime}}(-y) \,,
\label{eq:cs_upc}
\end{equation}
where $N_{\gamma/A}(y)$ is the photon flux; $y$ is the rapidity of the produced $\rho$ meson.
Because each ion can serve as a source of photons and a target, Eq.~(\ref{eq:cs_upc}) contains 
two contributions corresponding to the right-moving photon
source (first term) and the left-moving source (second term), respectively.
Equation~(\ref{eq:cs_upc}) implies the situation (experimental set-up), when the final state contains only 
a (reconstructed) $\rho$ meson, two large rapidity gaps and no special requirement 
is imposed on the number of forward nucleons, which are emitted in the 
nuclear break-up. However, requiring that UPCs are accompanied by forward 
neutron emission detected by zero-degree calorimeters (ZDCs) 
allows one to disentangle with a high probability and a
reasonable accuracy the two terms in Eq.~(\ref{eq:cs_upc}), see the discussion in Ref.~\cite{Guzey:2013jaa}.

The photon flux $N_{\gamma/A}(y)$ in Eq.~(\ref{eq:cs_upc}) is given by the convolution 
of the photon flux produced
by a fast moving ion at the distance $\vec{b}$ from its center,
 $N_{\gamma/A}(\omega,\vec{b})$~\cite{Vidovic:1992ik}, with the probability 
not to have strong interactions at given $\vec{b}$, $\Gamma_{AA}(b)$,
\begin{equation}
N_{\gamma/A}(y)=\int d^2 \vec{b}\, N_{\gamma/A}(\omega,\vec{b}) \Gamma_{AA}(\vec{b})  \,.
\label{eq:flux}
\end{equation}
where 
\begin{equation}
\Gamma_{AA}(\vec{b})=\exp\left(-\sigma_{NN} 
\int d^2 \vec{b}_1 T_A(\vec{b}_1) \, T_A(\vec{b}-\vec{b}_1) \right) \,.
\label{eq:Gamma_AA}
\end{equation}
In Eq.~(\ref{eq:Gamma_AA}), $T_A(\vec{b})=A\int^{\infty}_{-\infty} dz \rho_A(\vec{b},z)$ is the nuclear optical density,
where $\rho_A$ is the nuclear density, which we calculated using the Hartree--Fock--Skyrme 
approach~\cite{Beiner:1974gc}; $\sigma_{NN}$ is the energy-dependent nucleon--nucleon 
total cross section~\cite{Tanabashi:2018oca}.

Combining the vector meson dominance (VMD) model for the 
$\gamma N \to \rho N$ amplitude with 
the high-energy optical limit of the Glauber model 
and using the completeness (closure) of the nuclear final states $A^{\prime}$,
  the expression for the 
cross section of incoherent photoproduction
of $\rho$ mesons (and other vector mesons amenable to the VMD model) 
can be presented in the following form~\cite{Bauer:1977iq}
\begin{equation}
\sigma_{\gamma A \to \rho A^{\prime}}^{\rm Glauber}=
\left(\frac{e}{f_{\rho}}\right)^2 \int d^2 \vec{b}
\left[\langle 0|\Gamma_A^{\dagger}(b)\Gamma_A(b)|0 \rangle -\langle 0|\Gamma_A^{\dagger}(b)|0 \rangle
\langle 0|\Gamma_A(b)|0 \rangle
\right] \,,
\label{eq:sigG}
\end{equation}
where $f_{\rho}$ is the photon--meson coupling fixed by the $\rho \to e^{+} e^{-}$ decay width, 
$f_{\rho}^2/(4 \pi)=2.01 \pm 0.1$; the notation $\langle |\dots |\rangle $ 
stands for the integration with the ground-state
nuclear wave function squared (nuclear density).
In Eq.~(\ref{eq:sigG}), $\Gamma_A(b)$ is the $\rho$--nucleus scattering amplitude 
in the impact parameter space (profile function),
\begin{equation}
\Gamma_A(b)=1-\prod_{i=1}^{A}(1-\Gamma_N(b-s_i))  \,,
\label{eq:gamG}
\end{equation}
which is expressed through the 
$\rho$--nucleon 
amplitudes $\Gamma_N$,
\begin{equation}
\Gamma_N(b-s_i)=\frac{\sigma_{\rho N}}{4 \pi B} e^{-(b-s_i)^2/(2B)} \,,
\label{eq:gamN}
\end{equation}
where $s_i$ is the transverse coordinate of the $i$th nucleon; 
$\sigma_{\rho N}$ is the total $\rho$ meson--nucleon cross section;
$B $ is the slope of the $t$ dependence of the
 $\rho N \to \rho N$ cross section.
Note that in the high-energy limit, one can safely neglect 
the longitudinal momentum transfer to nucleons and the nucleon ordering.
Substituting Eqs.~(\ref{eq:gamG}) and (\ref{eq:gamN}) in Eq.~(\ref{eq:sigG}) 
and assuming independent nucleons in the nuclear wave function, 
one obtains
\begin{eqnarray}
&& \langle 0|\Gamma_A^{\dagger}(b)\Gamma_A(b)|0 \rangle-
\langle 0|\Gamma_A^{\dagger}(b)|0 \rangle \langle 0|\Gamma_A(b)|0 \rangle
= \left(1-\frac{\sigma_{\rho N}}{A}T_A(b)+\frac{\sigma_{\rho N}^2}{16\pi BA}T_A(b)
\right)^A-\left(1-\frac{\sigma_{\rho N}}{2A}T_A(b)\right)^{2A}\nonumber\\
& = & \exp\left[-\sigma_{\rho N} T_A(b)+\frac{\sigma_{\rho N}^2}{16\pi B}T_A(b)
\right]-\exp\left[-\sigma_{\rho N} T_A(b)\right] 
=\left(1-\exp\left[-\frac{\sigma_{\rho N}^2}{16\pi B}T_A(b)\right] \right) 
\exp\left[-\sigma_{\rho N} T_A(b)+\frac{\sigma_{\rho N}^2}{16\pi B}T_A(b)\right] \nonumber\\
& \approx  & 
\frac{\sigma_{\rho N}^2}{16\pi B}T_A(b)
\exp\left[-\left(\sigma_{\rho N}-\frac{\sigma_{\rho N}^2}{16\pi B}\right)T_A(b)
\right] \,.
\label{eq:glauber_b}
\end{eqnarray}
In the last line, we expanded in powers of the elastic 
$\rho$--nucleon cross section $\sigma_{\rho N}^{\rm el}=\sigma_{\rho N}^2/(16\pi B)$ 
and kept the leading contribution corresponding to 
the so-called one-step $\rho$ photoproduction process.
In the derivation of Eq.~(\ref{eq:glauber_b}), we used that the nuclear 
density $\rho_A(b,z)$ is a much slower function of the 
transverse coordinate $b$ than $\Gamma_N$, which allowed us to express 
the answer in a compact form in terms of $T_A(b)$.
Therefore, the final expression for the $\gamma A \to \rho A^{\prime}$ 
quasi-elastic incoherent cross section reads
\begin{equation}
\sigma_{\gamma A \to \rho A^{\prime}}^{\rm Glauber}=
\left(\frac{e}{f_{\rho}}\right)^2 \frac{\sigma_{\rho N}^2}{16\pi B}
\int d^2 \vec{b}\, T_A(b) e^{-\sigma_{\rho N}^{\rm in} T_A(b)} =
\sigma_{\gamma p \to \rho p} 
\int d^2 \vec{b}\, T_A(b) e^{-\sigma_{\rho N}^{\rm in} T_A(b)}  \,,
\label{eq:glauber}
\end{equation}
where 
$\sigma_{\rho N}^{\rm in}=\sigma_{\rho N}-\sigma_{\rho N}^{\rm el}$ is 
the inelastic $\rho$ meson--nucleon cross section; 
$\sigma_{\gamma p \to \rho p}=(e/f_{\rho})^2  \sigma_{\rho N}^{\rm el} $ is 
the elastic photoproduction 
cross section on the nucleon in the VMD model.  
Equation~(\ref{eq:glauber}) has a clear physical meaning and interpretation: 
photoproduction of $\rho$ mesons takes place 
on any of $A$ nucleons of the target, whose distribution in the transverse plane 
is given by $T_A(b)$, and the produced $\rho$ meson can further interact with the 
rest of target nucleons. While elastic interactions are allowed, 
the inelastic re-scattering would destroy
the final-state $\rho$ meson and, hence, should be rejected; 
the probability not to have inelastic scattering is given by 
$\exp[-\sigma_{\rho N}^{\rm in} T_A(b)]$.

Equation~(\ref{eq:glauber}) was derived assuming independent nucleons in 
the ground-state nuclear wave function. 
One can readily go beyond this approximation and take into account 
the effect of short-range nucleon--nucleon corrections (SRCs) in the nuclear 
wave function~\cite{Moniz:1971ih, VonBochmann:1969dn, Yennie:1986gm}. 
While the SRCs can noticeably modify the $t$ dependence of the incoherent cross section, 
the $t$-integrated cross section is influenced weakly.
Hence, this effect can be safely neglected in our analysis.

Equation~(\ref{eq:glauber}) implies that nuclear shadowing arises 
from rescattering of a single state with the
cross section $\sigma_{\rho N}$, i.e., that the effect of diffractively 
produced states leading to the inelastic (Gribov)
 correction is neglected. 
A convenient way to take into account the inelastic shadowing correction is 
offered by the formalism of cross section fluctuations capturing the composite 
hadronic structure of the photon, see, e.g.~\cite{Frankfurt:2015cwa,Guzey:2016piu,Guzey:2013jaa}.
In this approach~\cite{Good:1960ba,Miettinen:1978jb,Kopeliovich:1981pz,Blaettel:1993ah}, 
the key quantity is the distribution $P(\sigma)$ giving the probability 
for the hadronic component of the photon to interact with the nucleon with the cross section $\sigma$. 
Following the analysis of Refs.~\cite{Frankfurt:2015cwa,Guzey:2016piu}, 
we parametrize this distribution in the
following form
\begin{equation}
P(\sigma)=\frac{C}{1+(\sigma/\sigma_0)^2}e^{-[\sigma/\sigma_0)^2-1]^2/\Omega^2} \,.
\label{eq:P}
\end{equation}
The free parameters $C$, $\sigma_0$, and $\Omega$ are found from the constraints 
on the first three moments of the distribution
$P(\sigma)$:
\begin{eqnarray}
\int_0^{\infty} d\sigma P(\sigma) &=& 1 \,, \nonumber\\
\int_0^{\infty} d\sigma P(\sigma)\sigma &=& \hat{\sigma}_{\rho N}(W_{\gamma p})\,, \nonumber\\
\int_0^{\infty} d\sigma P(\sigma) \sigma^2 &=& [\hat{\sigma}_{\rho N}(W_{\gamma p})]^2 \left(1+\omega_{\sigma}(W_{\gamma p})\right) \,.
\label{eq:P2}
\end{eqnarray}
The first equation is the probability conservation. The second equation constrains the average value of $P(\sigma)$ and
implies that the hadronic fluctuations of the photon
should lead to the effective $\rho$--nucleon cross section $\hat{\sigma}_{\rho N}$.
As discussed in Ref.~\cite{Frankfurt:2015cwa}, the $\hat{\sigma}_{\rho N}$ effective cross section 
determined from the fit to the available data on the forward 
$d\sigma_{\gamma p \to \rho p}(t=0)/dt$ cross section using
the VMD relation,
\begin{equation}
\hat{\sigma}_{\rho N}(W_{\gamma p})=\left(\frac{f_{\rho}^2}{4 \pi \alpha_{\rm em}} 16 \pi d\sigma_{\gamma p \to \rho p}(W_{\gamma p},t=0)/dt \right)^{1/2} \,,
\label{eq:P3}
\end{equation}
turns out to be somewhat smaller than an estimate based on the constituent quark model because of an enhanced
contribution of small-$\sigma$ fluctuations. It is reflected in the form of $P(\sigma)$ in Eq.~(\ref{eq:P}) and generally
leads to a violation of the naive VMD model. 
The third equation of Eqs.~(\ref{eq:P2}) constrains the dispersion of $P(\sigma)$, which is parameterized by 
$\omega_{\sigma}$. In our analysis, following Ref.~\cite{Frankfurt:2015cwa}, we
 relate it to the corresponding parameter for the pion and 
 use $\omega_{\sigma}=0.3 \pm 0.05$.
 This uncertainty in $\omega_{\sigma}$ leads to the uncertainty of our predictions of the nuclear cross sections, which we show
 by red shaded bands in Figs.~\ref{fig:cs_incoh_UPC} and \ref{fig:cs_coh} below.

In incoherent $\rho$ photoproduction on nuclei, hadronic fluctuations of the photon act at the level of the $\gamma A \to \rho A^{\prime}$ amplitude. The corresponding 
quasi-elastic incoherent cross section can be readily obtained
by generalizing the derivation of Eq.~(\ref{eq:glauber}) [the superscript ``GG'' stands for Gribov--Glauber]
\begin{eqnarray}
\sigma_{\gamma A \to \rho A^{\prime}}^{\rm GG} &=&
\left(\frac{e}{f_{\rho}}\right)^2 \int d^2 \vec{b} \int d\sigma P(\sigma)\int d\sigma^{\prime} P(\sigma^{\prime}) 
\frac{\sigma \sigma^{\prime}}{16\pi B}T_A(b)\exp\left[-\frac{\sigma+\sigma^{\prime}}{2}T_A(b)+\frac{\sigma \sigma^{\prime}}{16\pi B}T_A(b)\right] \nonumber\\
&=& \left(\frac{e}{f_{\rho}}\right)^2 \int d^2 \vec{b}\, T_A(b) \left(\int d\sigma P(\sigma) 
\frac{\sigma}{\sqrt{16\pi B}}\exp\left[-\frac{\sigma^{\rm in}}{2}T_A(b)\right]\right)^2\,.
\label{eq:inc7}
\end{eqnarray}
To present the answer in a compact form,  in the exponential factor in the first line we used that 
$\sigma \sigma^{\prime}/(16\pi B) = [\sigma^2/(16\pi B)+\sigma^{\prime 2}/(16\pi B)]/2-(\sigma-\sigma^{\prime})^2/(32\pi B)$
and neglected the contribution of the second term, whose contribution is small because both $\sigma$ and $\sigma^{\prime}$ are distributed (fluctuate) around the same average cross section.
Neglecting hadronic fluctuations of the photon, i.e., replacing $P(\sigma)$ by the  $\delta$-function in Eq.~(\ref{eq:inc7}),
one obtains the Glauber model expression of Eq.~(\ref{eq:glauber}).

For comparison and completeness, we also give the cross section of coherent $\rho$ photoproduction in the same approach~\cite{Frankfurt:2015cwa}
\begin{equation}
\sigma_{\gamma A \to \rho A}^{\rm GG}=\left(\frac{e}{f_{\rho}}\right)^2 \int d^2 \vec{b} \left|\int d\sigma P(\sigma) \left(1-e^{-\frac{\sigma}{2}T_A(\vec{b})}\right) \right|^2 \,.
\label{eq:cs_GG}
\end{equation}
In the absence of the fluctuations, i.e., using $P(\sigma)=\delta(\sigma-\sigma_{\rho N})$ in Eq.~(\ref{eq:cs_GG}), 
one obtains the standard expression for the coherent $\rho$ photoproduction cross section in the optical limit of the 
Glauber model.

\section{Results, discussion and comparison to STARlight}
\label{sec:results}

As follows from the final expressions for the cross section of incoherent $\rho$ photoproduction on heavy nuclei, see 
Eqs.~(\ref{eq:glauber}) and (\ref{eq:inc7}),
predictions for the nuclear cross section directly depend either on the $t$-integrated $\sigma_{\gamma p \to \rho p}(W_{\gamma p})$ 
cross section on the proton or on the differential $d\sigma_{\gamma p \to \rho p}(W_{\gamma p},t=0)/dt$ cross section extrapolated to
$t=0$; additionally, the effect of nuclear attenuation (shadowing) depends on 
the slope $B$ of the $t$ dependence of the $\gamma p \to \rho p$ cross section.
In our analysis, we calculate the photoproduction cross section on the proton,
\begin{eqnarray}
\frac{d\sigma_{\gamma p \to \rho p}(W_{\gamma p},t)}{dt}=\left|T_{\gamma p\rightarrow \rho p}\right|^2 \,,
\end{eqnarray}
using the Donnachie--Landshoff (DL) model~\cite{Donnachie:1999yb,Donnachie:2008kd} 
accounting for contributions of the soft DL Pomeron (P), Reggeon (R), and hard
pomeron (H) exchanges
in the amplitude $T_{\gamma p\rightarrow \rho p}$,
\begin{eqnarray}
T_{\gamma p\rightarrow \rho p}(s,t) = 
iF_{p}(t)G_{\rho}(t)\bigl [C_{P}e^{-{\frac {1} {2}}i\pi \alpha_{P}(t)}
(2\alpha_{P}^{\prime}s)^{\alpha_{P} (t)-1} + 
C_{R}e^{-{\frac {1} {2}}i\pi \alpha_{R}(t)}
( 2\alpha_{R}^{\prime}s)^{\alpha_{R} (t)-1} \bigr ]+T_{H}(s,t) \,,
\label{eq:elem}
\end{eqnarray}
where $F_1^p(t)$ and $G_{\rho}(t)$ are the proton Dirac and the $\gamma \rho$ vertex 
form factors; $\alpha_P(t)=\alpha_P(0)+\alpha_P^{\prime}t$ 
is the soft Pomeron trajectory; 
$\alpha_R(t)=\alpha_R(0)+\alpha_R^{\prime}t$ 
is the Reggeon trajectory;
$s=W_{\gamma p}^2$.
 In our calculations, we have used the values of $\alpha_P(0)=1.093$, 
 $\alpha_P^{\prime}=0.25$ GeV$^{-2}$ and
$\alpha_R(0)=0.5$, $\alpha_R^{\prime}=0.93$ GeV$^{-2}$ for the intercepts and the slopes
 of the soft Pomeron and Reggeon trajectories, respectively. Also, we slightly readjusted the
 values of the normalization constants $C_P$ and $C_R$ to describe better the
 energy dependence of the experimental total and forward cross sections  of the 
$\gamma p\rightarrow \rho p$ process measured in a wide range of energies.
The parameters fixing the contribution of the hard Pomeron exchange were taken from~\cite{Donnachie:1999yb,Donnachie:2008kd}.
Note that the DL hard Pomeron gives a significant contribution to 
$d\sigma_{\gamma p \to \rho p}(W_{\gamma p},t)/dt$ only for large $-t>1$ GeV$^2$
and at energies much higher than the considered range. 
Because
our calculation of the $\gamma p \to \rho p$ photoproduction cross section relies on
the vector meson dominance (VMD) and the use of Eq.~(\ref{eq:elem}),
we refer to this approach as the modified 
DL  (mDL) model.

\begin{figure}[t]
\begin{center}
\vspace{-3cm}
\epsfig{file=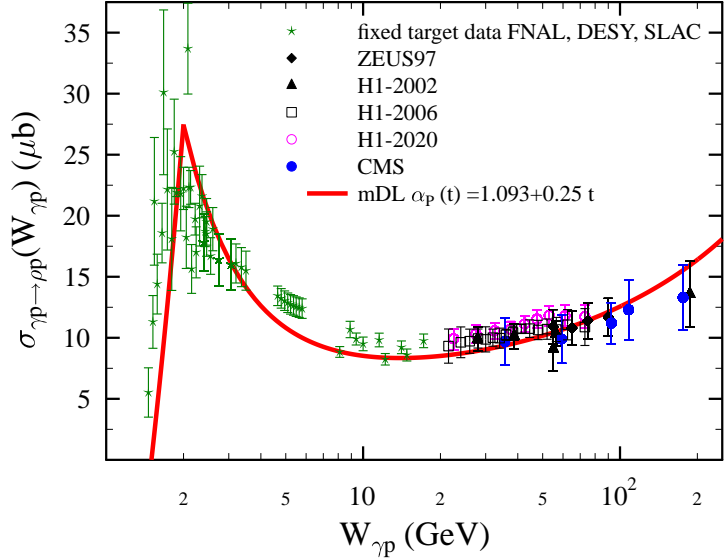,scale=0.6}
\vspace{-1cm}
\caption{The $\sigma_{\gamma p \to \rho p}$ cross section as a function of the 
photon--proton energy $W_{\gamma p}$: the 
mDL model
 vs.~the available fixed-target, HERA, and $pA$ UPC at the LHC data.}
\label{fig:elem1}
\end{center}
\end{figure}

\begin{figure}[h]
\begin{center}
\vspace{-2.5cm}
\epsfig{file=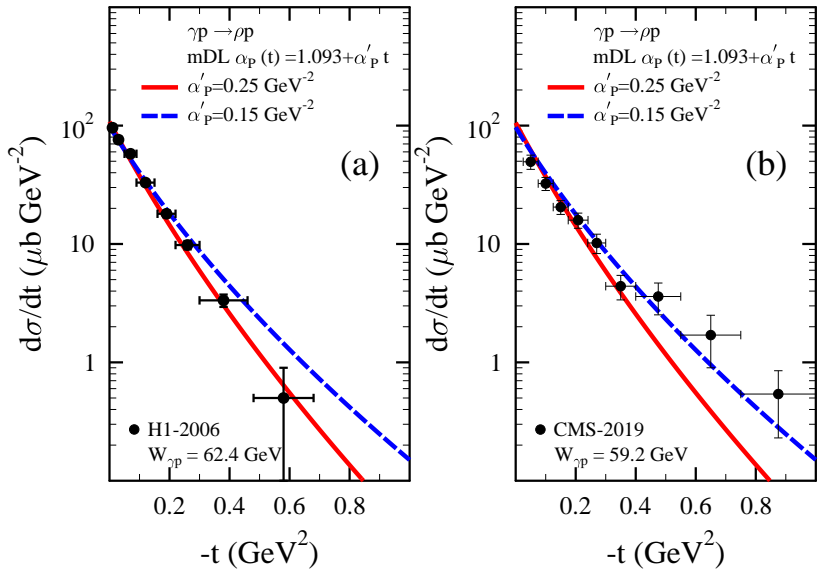,scale=0.6}
\vspace{-1cm}
\caption{The $d\sigma_{\gamma p \to \rho p}(W_{\gamma p})/dt$ differential cross section in the mDL model with
$\alpha_P^{\prime}=0.15$ GeV$^{-2}$ and $\alpha_P^{\prime}=0.25$ GeV$^{-2}$ 
values of the slope the Pomeron trajectory 
vs.~the H1 data~\cite{Weber:2006di} at $W_{\gamma p}=62.4$ GeV (panel (a)) and
the CMS data~\cite{Sirunyan:2019nog} at $W_{\gamma p}=59.2$ GeV (panel (b)).}
\label{fig:elem2}
\end{center}
\end{figure}

Figure~\ref{fig:elem1} shows the $\sigma_{\gamma p \to \rho p}$ cross 
section as a function of the 
photon--proton energy $W_{\gamma p}$: over a broad
range of $W_{\gamma p}$, our mDL model given by the solid curve describes 
well the available data obtained with
fixed targets (SLAC~\cite{Park:1971ts}, CERN~\cite{Aston:1982hr}, FNAL~\cite{Egloff:1979mg}), at the HERA lepton--proton 
collider~\cite{Breitweg:1997ed,Weber:2006di,H1:2020lzc}, 
and in the proton--nucleus ($pA$) ultraperipheral collisions (UPCs) 
by the CMS collaboration at the LHC~\cite{Sirunyan:2019nog}.
Note that the theoretical cross section was linearly extrapolated to the threshold for $W_{\gamma p} < 2$ GeV.
It is important to point out that the analysis of the 2020 H1 data gives 
$\alpha_P^{\prime}=0.233^{+0.067}_{-0.074}$ GeV$^{-2}$~\cite{H1:2020lzc}, which agrees with the value used in the mDL model~(\ref{eq:elem}) employed in our calculations.

The $d\sigma_{\gamma p \to \rho p}(W_{\gamma p})/dt$ differential cross section in the mDL model with
$\alpha_P^{\prime}=0.15$ GeV$^{-2}$ and $\alpha_P^{\prime}=0.25$ GeV$^{-2}$ 
values of the slope the Pomeron trajectory is shown in Fig.~\ref{fig:elem2}. The results of the model are compared to
the H1 data~\cite{Weber:2006di} at $W_{\gamma p}=62.4$ GeV (panel (a)) and
the CMS data~\cite{Sirunyan:2019nog} at $W_{\gamma p}=59.2$ GeV (panel (b)). These values 
 of $W_{\gamma p}$ correspond to the photon--nucleon energy 
in $\rho$ photoproduction at $y=0$ in Pb-Pb UPCs at $\sqrt{s_{NN}}=5.02$ TeV.
One can see from this figure that while the H1 data favors a steeper
$|t|$ dependence with the $\alpha^{\prime}_P=0.25$ GeV$^{-2}$ slope of the Pomeron trajectory, the recent 
CMS data~\cite{Sirunyan:2019nog} seems to prefer the lower
value of the slope,  $\alpha^{\prime}_P=0.15$ GeV$^{-2}$.
This illustrates the current uncertainty in the value of 
the $d\sigma_{\gamma p \to \rho p}(W_{\gamma p},t)/dt$ cross section 
serving as input for the calculation of the cross section of incoherent $\rho$ photoproduction on nuclei.

\begin{figure}[t]
\begin{center}
\epsfig{file=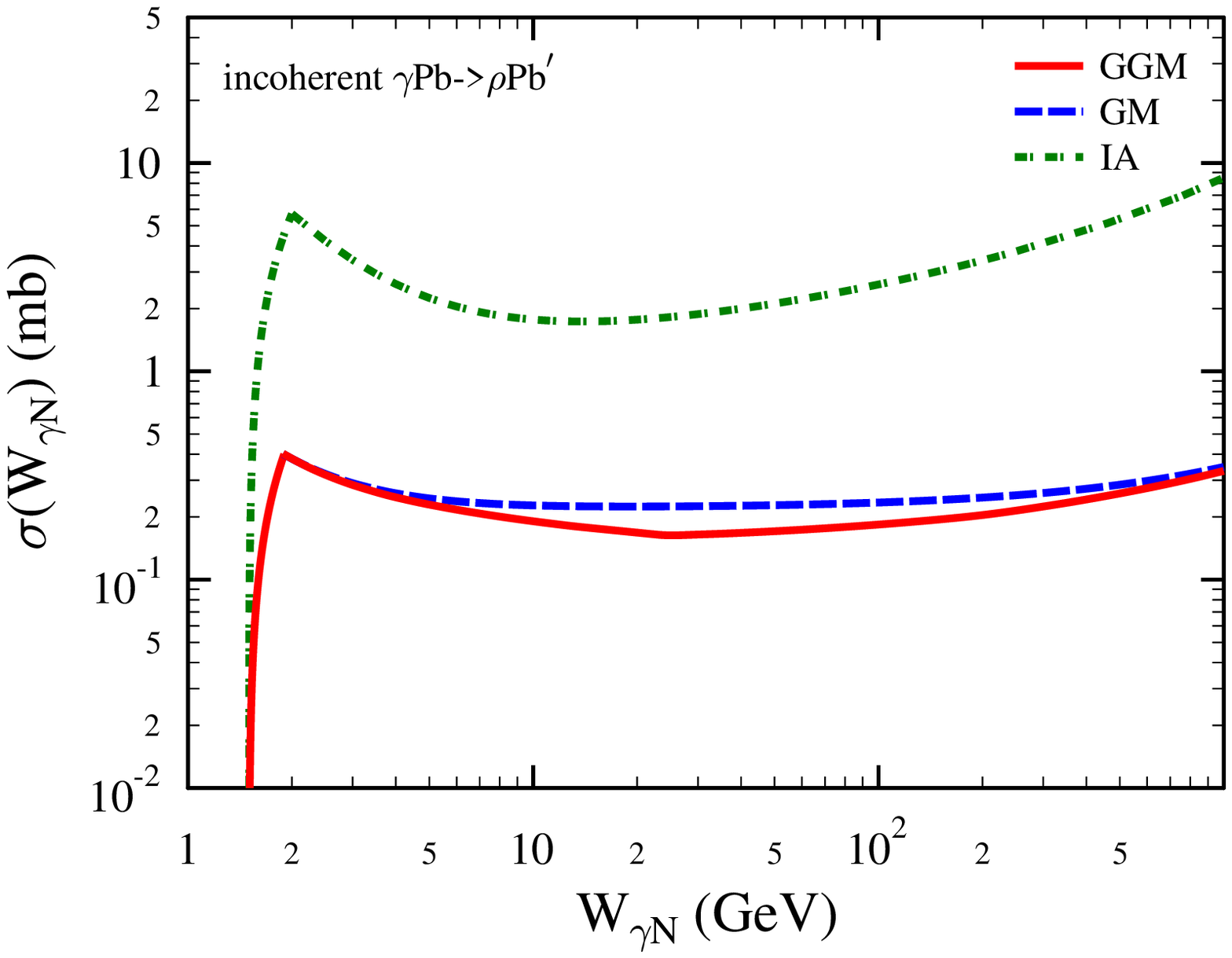,scale=0.6}
\vspace{-1cm}
\caption{The $\sigma_{\gamma A \to \rho A^{\prime}}(W_{\gamma p})$ 
incoherent cross section as a function of 
$W_{\gamma p}$ in the Gribov--Glauber model (red solid curve labeled ``GGM"), 
the Glauber model (blue dashed curve labeled ``GM"), 
and the impulse approximation (green dot-dashed curve).}
\label{fig:cs_incoh}
\end{center}
 \end{figure}

\begin{figure}[t]
\begin{center}
\vspace{-2cm}
\epsfig{file=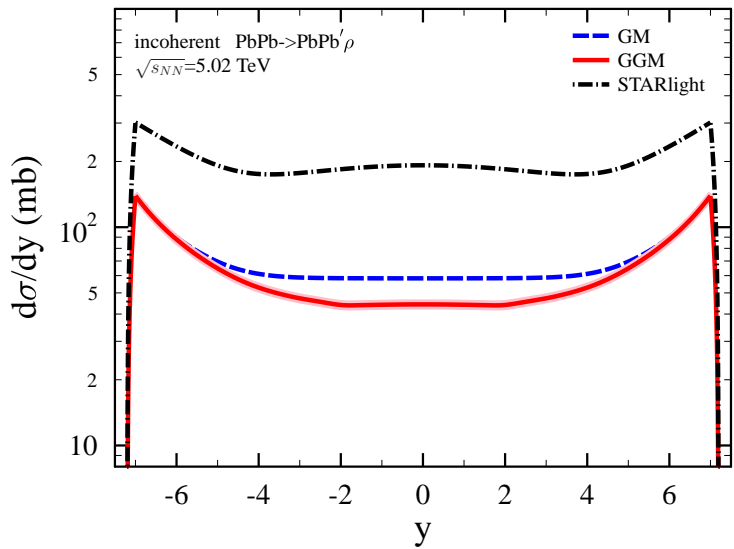,scale=0.6}
\caption{The incoherent UPC cross section $d \sigma_{AA \to \rho AA^{\prime}}/dy$
as a function of the $\rho$ meson rapidity $y$ at $\sqrt{s_{NN}}=5.02$ TeV. 
Shown are predictions of the Gribov--Glauber (red solid curve with a shaded band),
Glauber (blue dashed),
and STARlight (black dot-dashed curve) models.}
\label{fig:cs_incoh_UPC}
\end{center}
 \end{figure}

\begin{figure}[t]
\begin{center}
\vspace{-2.5cm}
\epsfig{file=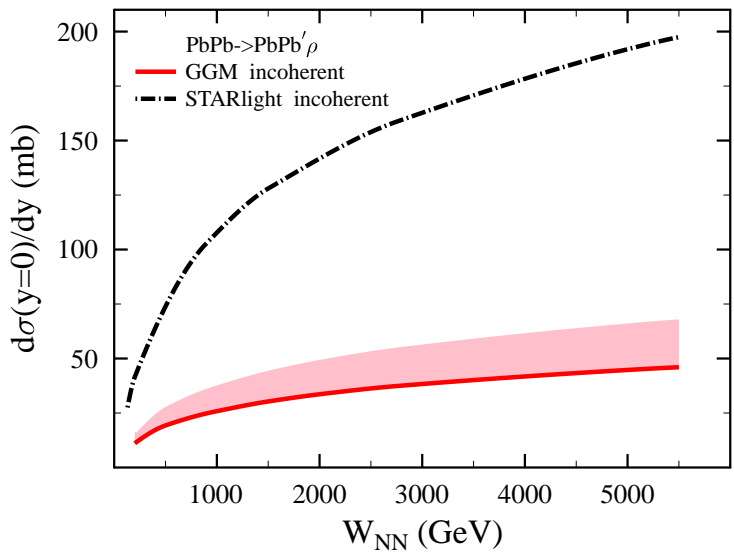,scale=0.6}
\vspace{-0.5cm}
\caption{The incoherent UPC cross sections as a function 
of $W_{NN}=\sqrt{s_{NN}}$ at $y=0$ in the Gribov--Glauber (red solid) 
and STARlight (black dot-dashed) models.
The red shaded band shows the range of predictions for the cross section of incoherent 
$\rho$ photoproduction on nuclei, which includes both elastic and nucleon-dissociative photoproduction on target nucleons,
see Eq.~(\ref{eq:inc7_dd}).
}
\label{fig:cs_energy}
\end{center}
\end{figure}

\begin{figure}[t]
\begin{center}
\vspace{-2cm}
\epsfig{file=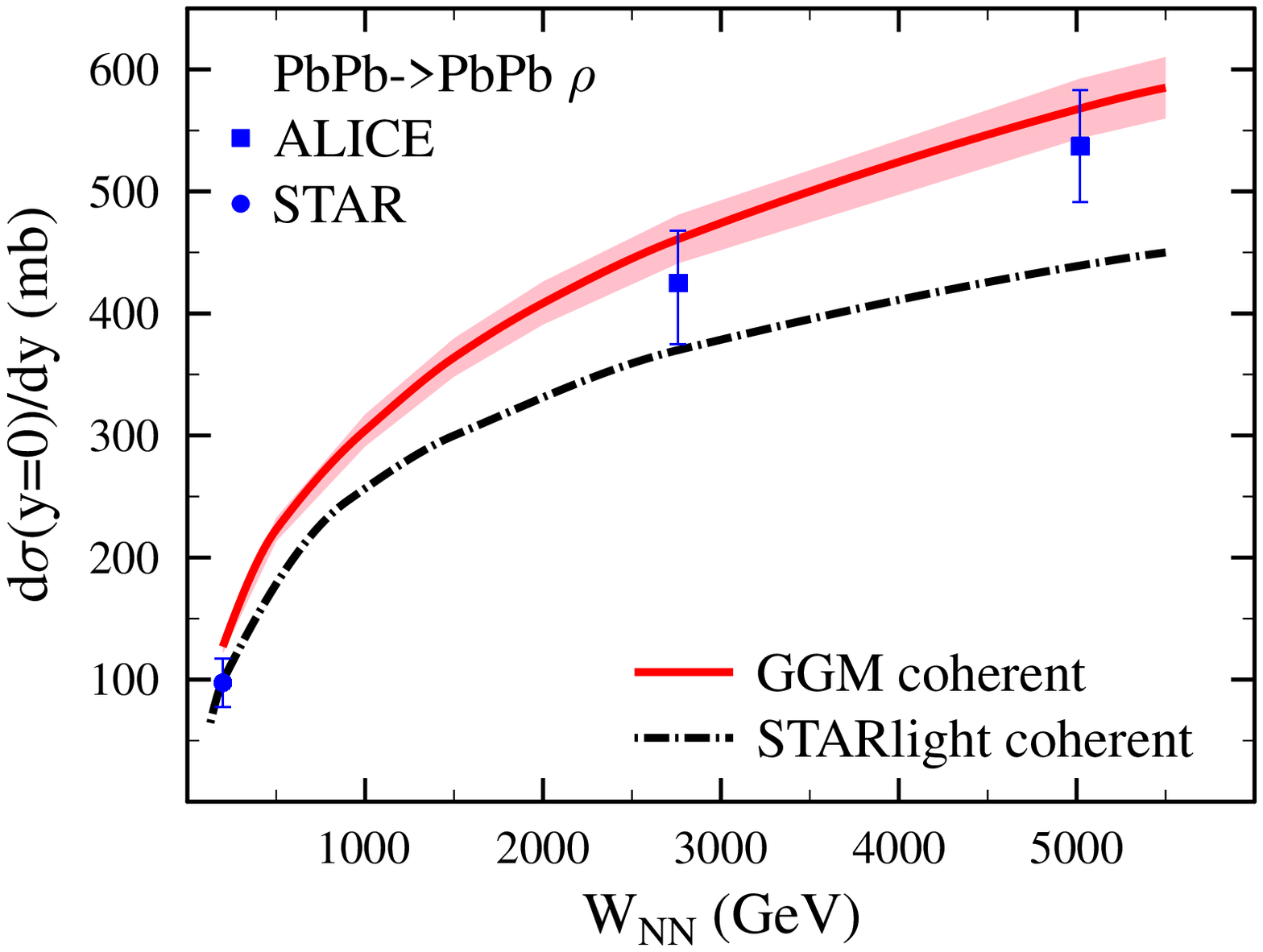,scale=0.6}
\vspace{-1cm}
\caption{The coherent UPC cross sections as a function 
of $W_{NN}=\sqrt{s_{NN}}$ at $y=0$ in the Gribov--Glauber (red solid curve with a shaded band) 
and STARlight (black dot-dashed curve) models.
The scaled STAR measurement at $\sqrt{s_{NN}}=200$ GeV~\cite{Abelev:2007nb} and
the ALICE measurements at $\sqrt{s_{NN}}=2.76$ TeV~\cite{Adam:2015gsa} and $\sqrt{s_{NN}}=5.02$ TeV~\cite{Acharya:2020sbc}
 are shown by the filled circle and the squares with error bars, respectively.
}
\label{fig:cs_coh}
\end{center}
\end{figure}

Our predictions for the cross sections of $\rho$ photoproduction 
in nucleus--nucleus UPCs in the LHC kinematics are
presented in Figs.~\ref{fig:cs_incoh}, \ref{fig:cs_incoh_UPC}, \ref{fig:cs_energy}, and \ref{fig:cs_coh}.

 Figure~\ref{fig:cs_incoh} shows the $\gamma A \to \rho A^{\prime}$ 
incoherent cross section as a function of the 
photon--nucleon energy $W_{\gamma p}$. The three curves correspond 
to the results of the calculations using the Gribov--Glauber model of Eq.~(\ref{eq:inc7}) 
(the red solid curve labeled ``GGM"), the Glauber model of Eq.~(\ref{eq:glauber})
(the blue dashed curve labeled ``GM"), and the impulse approximation 
(the green dot-dashed curve), where one neglects the effect 
of nuclear attenuation, $\sigma_{\gamma A \to \rho A^{\prime}}^{\rm IA}=A \sigma_{\gamma p \to \rho p}$. 
One can see from the figure that the effect of nuclear shadowing is 
very large and leads to the suppression of the incoherent
cross section compared to the impulse approximation by the factor of ten. Also,  one can see that the 
inelastic nuclear shadowing additionally reduces
the incoherent cross section by about 25\% in a broad range of $W_{\gamma p}$.
Note that in our calculations in the Gribov--Glauber model, 
in the interval $20 \leq W_{\gamma p} \leq 200$ GeV we used the nominal central value for the parameter 
quantifying the photon hadronic fluctuations, 
 $\omega_{\sigma}=0.3$. 
 For $W_{\gamma p} > 200$ GeV, based on data on inelastic diffraction in antiproton-proton and proton-proton scattering at Tevatron and  LHC energies, it is expected that $\omega_{\sigma}$ decreases with an increase of energy
 and eventually vanishes at asymptotically high energies because of an onset of the so called black disk limit~\cite{Frankfurt:2015cwa}. 
 This results in a gradual decrease of inelastic nuclear shadowing for very large $W_{\gamma p}$ leading to
 convergence of the red solid and blue dashed curves. 
 The uncertainty in $\omega_{\sigma}$ leads to a small, of the order of 5\%, uncertainty in the predicted
 incoherent cross section, which is significantly smaller than the difference among the three curves shown in this figure.

Figure~\ref{fig:cs_incoh_UPC} shows the incoherent UPC cross section~(\ref{eq:cs_upc}) as a function of the $\rho$ meson rapidity 
$y$ at $\sqrt{s_{NN}}=5.02$ TeV. The red solid curve corresponds to
$\sigma_{\gamma A \to \rho A^{\prime}}$ in the Gribov--Glauber model;
the blue dashed curve is the result of the Glauber model, c.f.~Fig.~\ref{fig:cs_incoh}
and its discussion.
The shaded band shows the theoretical uncertainty of our predictions due to the 
 uncertainty in the value of $\omega_{\sigma}$, which we take to be $\omega_{\sigma}=0.3 \pm 0.05$~\cite{Frankfurt:2015cwa}.
The black dot-dashed 
curve is the result of 
the STARlight model~\cite{Klein:2016yzr}. 
One can see that over essentially the entire range of $y$, 
the predictions of the Gribov--Glauber model lie dramatically lower than those of the  STARlight model.

In the STARlight framework, it is assumed that the cross section of incoherent 
photoproduction of vector mesons on nuclear
targets is proportional to the ratio of
the inelastic $\rho A$ and $\rho N$ cross sections
and is given by the following expression
\begin{equation}
\sigma_{\gamma A \to \rho A^{\prime}}^{\rm STARlight}=\frac 
{\sigma_{\gamma p \to \rho p}} {{\sigma_{\rho N}^{\rm in}}} 
{\sigma_{\rho A}^{\rm in}} =
\left(\frac{e}{f_{\rho}}\right)^2 \frac{\sigma_{\rho N}^{\rm el}}{\sigma_{\rho N}^{\rm in}}
\sigma_{\rho A}^{\rm in} \,,
\label{eq:sl1}
\end{equation}
where $\sigma_{\rho A}^{\rm in}$ is calculated in the Glauber model,
\begin{equation}
\sigma_{\rho A}^{\rm in}=\int d^2 \vec{b} \left(1-e^{-\sigma_{\rho N}T_A(b)} \right) \,.
\label{eq:sl2}
\end{equation}
Equation~(\ref{eq:sl1}) does not 
correspond to the Glauber expression for the  
quasi-elastic
incoherent $\gamma A \to \rho A^{\prime}$ cross section. 
As follows from unitary of the Glauber theory, the inelastic $\rho A$ cross section 
can be presented by 
a sum of partial cross sections 
of the inelastic interactions of the produced $\rho$-meson
with $N_{\rm eff}$ nucleons ($ 1 \leq N_{\rm eff} \leq A$)~\cite{Bertocchi:1976bq}, 
which lead to the
final state with a much higher multiplicity than that in
the incoherent cross section studied in UPCs.
As a result, 
the STARlight predictions for the 
cross section of incoherent photoproduction of $\rho$ mesons in Pb-Pb UPCs 
at $\sqrt{s_{NN}}= 5.02$ TeV significantly overestimate those obtained in 
our GGM approach, see also Table~\ref{table:sl}.

It is also illustrated by Fig.~\ref{fig:cs_energy}, which shows the incoherent 
$d \sigma_{AA \to \rho AA^{\prime}}/dy$ UPC cross sections as 
a function of the center-of-mass energy
$W_{NN}=\sqrt{s_{NN}}$ at the central rapidity $y=0$ and 
compares the results of the Gribov--Glauber and STARlight models.
One can see from this figure that the STARlight predictions are several-fold larger than those of the
Gribov--Glauber model.

The quasi-elastic incoherent cross sections of Eqs.~(\ref{eq:glauber}) 
and (\ref{eq:inc7}) do not include the contribution of 
$\rho$ photoproduction with nucleon dissociation, $\gamma N \to \rho Y$, 
where $Y$ denotes the hadronic system with mass $M_Y$. 
If this contribution is not rejected experimentally, it will
increase the incoherent cross section. The effect can be taken into 
account using the approach developed for 
incoherent $J/\psi$ photoproduction on nuclei~\cite{Guzey:2018tlk}. 
Using this method, the cross section of incoherent 
$\rho$ photoproduction on nuclei, which includes both elastic and 
nucleon-dissociative photoproduction on target nucleons, can be presented 
in the following form [compare to Eq.~(\ref{eq:inc7})]
\begin{equation}
\sigma_{\gamma A \to \rho A^{\prime}+Y}^{\rm GG} =\left(1+\frac{\sigma_{\gamma p \to \rho Y}}{\sigma_{\gamma p \to \rho p}}\right)
 \left(\frac{e}{f_{\rho}}\right)^2 \int d^2 \vec{b}\, T_A(b) \left(\int d\sigma P(\sigma) 
\frac{\sigma}{\sqrt{16\pi B}}\exp\left[-\frac{\sigma^{\rm in}}{2}T_A(b)\right]\right)^2 \,,
\label{eq:inc7_dd}
\end{equation}
where $\sigma_{\gamma p \to \rho p}$ and $\sigma_{\gamma p \to \rho Y}$ are 
the $t$-integrated cross sections of 
elastic and nucleon-dissociative $\rho$ photoproduction on the proton, respectively.
Using the ZEUS analysis of elastic and proton-dissociative $\rho^0$ photoproduction at HERA~\cite{Breitweg:1997ed}
that found
$\sigma_{\gamma p \to \rho p}/\sigma_{\gamma p \to \rho Y}=2.0 \pm 0.2 ({\rm stat}.) \pm 0.7 (\rm syst.)$
in kinematic domain $M_Y < 0.1W_{\gamma p}^2$ and $|t|<0.5$ GeV$^2$,
we estimate that
the nucleon dissociation may increase the cross section of one-step
incoherent $\rho$ photoproduction by as much as 50\%.
The exact magnitude of this contribution depends on such data selection 
criteria as the mass of the produced state $Y$ and
the range of the momentum transfer $t$. 
To reflect it, the possible range of our predictions 
for $\sigma_{\gamma A \to \rho A^{\prime}+Y}^{\rm GG}$ of Eq.~(\ref{eq:inc7_dd}) 
is given by the red shaded band in Fig.~\ref{fig:cs_energy}.
Note that the uncertainty in the value of $\omega_{\sigma}$ 
results in a small, 5\% uncertainty in the predicted incoherent cross section, which can be neglected compared
to the magnitude of the nucleon-dissociation contribution.

Note that predictions of the Glauber model fall within the range of the shaded band and, hence, are not shown;
the difference between the Gribov--Glauber and Glauber model predictions can be readily read off 
Fig.~\ref{fig:cs_incoh_UPC}.

Predictions for the cross section of incoherent $\rho$ photoproduction in Pb-Pb UPCs in the LHC kinematics 
including the effect of nucleon dissociation were also made in the hot-spot model in Ref.~\cite{Cepila:2018zky}. While
the relative magnitude of the proton-dissociative contribution to the incoherent cross section is similar to our result, 
the absolute value of the incoherent cross section is several times smaller than our estimate. Thus, future
measurements of incoherent $\rho$ photoproduction in heavy-ion UPCs at the LHC
will help to discriminate between the discussed approaches and constrain the dynamics of nuclear shadowing
in vector meson photoproduction on nuclei.

It is also instructive to compare the incoherent and coherent cases.
 The cross section of coherent 
$\rho$ photoproduction on nuclei in the STARlight model 
is given by the following expression
\begin{equation}
\sigma_{\gamma A \to \rho A}^{\rm STARlight}=\frac{d\sigma_{\gamma A \to \rho A}(t=0)}{dt} 
\int_{|t_{\rm min}|}^{\infty} dt F_A^2(t)=
\left(\frac{e}{f_{\rho}}\right)^2 \frac{\sigma_{\rho A}^2}{16\pi} 
\int_{|t_{\rm min}|}^{\infty} dt F_A^2(t)\,,
\label{eq:sl2_coh}
\end{equation}
where $|t_{\rm min}|=(M_{\rho}^2 m_N/W_{\gamma p})^2$.
The factorized form of the
expression in Eq.~(\ref{eq:sl2_coh}) [compare to Eq.~(\ref{eq:cs_GG}) in the Gribov--Glauber model]
assumes that 
the $t$ dependence of the amplitude of coherent photoproduction on nuclei 
can be approximated by the undistorted nuclear form factor $F_A(t)$. 
This assumption disagrees with the Glauber model, see, e.g.~\cite{Frankfurt:2002sv},
 and does not only result in the $t$ dependence of the
 $d \sigma_{AA \to \rho AA}/(dy \, dt)$ cross section, which is wider than that predicted
in the Gribov--Glauber approach~\cite{Frankfurt:2015cwa}
 and seemed to be observed in the data~\cite{Adam:2015gsa},
but also increases the $t$ integrated cross section by a factor about $1.3 - 1.4$.
At the same time, in the standard option of the STARlight model, the $\sigma_{\rho A}$ cross section in
Eq.~(\ref{eq:sl2_coh}) is identified with the inelastic $\sigma^{\rm in}_{\rho A}$ cross section~(\ref{eq:sl2})
instead of the total one. It leads to a strong energy-dependent violation of the optical theorem and 
a suppression of the coherent cross section by approximately a factor of three
in the LHC kinematics (a factor of four in the asymptotic black body limit). 
An interplay of these two effects results in an overall suppression by approximately a factor of two 
of the cross section of coherent $\rho$ photoproduction on nuclei in the STARlight model compared to 
the standard optical-limit Glauber model, see Table~\ref{table:sl}.
Note that the STARlight framework has an option for the calculation of the $\sigma_{\gamma A \to \rho A}^{\rm STARlight}$ cross
section with the total $\rho$--nucleus cross section 
calculated in the Glauber model. It leads
to a very large value of the coherent cross section at $\sqrt{s_{NN}}= 5.02$ TeV,
$d \sigma^{\rm STARlight}_{AA \to \rho AA}/dy(y=0) \approx 1100$ mb.

Predictions of the Gribov--Glauber and STARlight models 
for the coherent $d \sigma_{AA \to \rho AA}/dy$ UPC cross section as 
a function of $W_{NN}=\sqrt{s_{NN}}$ at $y=0$ are shown in Fig.~\ref{fig:cs_coh}.
Also, the scaled results of the STAR measurement of coherent $\rho$ photoproduction 
in Au-Au UPCs at $\sqrt{s_{NN}}=200$ GeV~\cite{Abelev:2007nb} and
ALICE measurements of coherent $\rho$ photoproduction 
in Pb-Pb UPCs at $\sqrt{s_{NN}}=2.76$ TeV~\cite{Adam:2015gsa} 
and $\sqrt{s_{NN}}=5.02$ TeV~\cite{Acharya:2020sbc}
are shown by the filled circle and the squares with error bars, respectively.
One can see that the predictions of our approach are in excellent agreement with the ALICE data.
Note that the STAR data point for Au was scaled to Pb by the ratio of the theoretical cross sections.
The Glauber model prediction (not shown) significantly exceeds that of the Gribov--Glauber approach and, hence, fails to describe
the Run 1 and 2 ALICE data points, see Ref.~\cite{Frankfurt:2015cwa}
 and Table~\ref{table:sl} of the present work.

Table~\ref{table:sl} summarizes the results for the incoherent 
$d \sigma_{AA \to \rho AA^{\prime}}/dy$ and coherent $d \sigma_{AA \to \rho AA}/dy$ 
cross sections of $\rho$ photoproduction
in Pb-Pb UPCs at $\sqrt{s_{NN}}= 5.02$ TeV and $y=0$ in the framework presented 
in this paper (GM and GGM)
and the STARlight model. 
It clearly demonstrates large differences between predictions of the Gribov--Glauber model superseding the Glauber model and
those of 
STARlight, which are especially dramatic for the incoherent cross section.

\begin{table}[h]
\caption{Predictions for the incoherent $d \sigma_{AA \to \rho AA^{\prime}}/dy$ 
and coherent $d \sigma_{AA \to \rho AA}/dy$ cross sections (in mb) of $\rho$ photoproduction
in Pb-Pb UPCs at $\sqrt{s_{NN}}= 5.02$ TeV and $y=0$ in the framework presented 
in this paper (GM and GGM) and the STARlight model.}
\begin{center}
\begin{tabular}{|c|c|c|c|}
\hline 
& GM & GGM & STARlight \\
\hline
Incoherent, mb & 58 & 44 &  192 \\
Coherent, mb   & 840 & 570 & 440 \\
\hline
\end{tabular}
\end{center}
\label{table:sl}
\end{table}%

\section{Conclusions}
\label{sec:summary}

In this paper, using the Gribov--Glauber model for photon--nucleus scattering and a 
generalization of the VMD model 
for the hadronic structure of the photon,  
we consider incoherent photoproduction of $\rho$ mesons on heavy nuclei and make predictions for 
the incoherent $Pb Pb \to \rho Pb A^{\prime}$ UPC cross section in the LHC kinematics.
We present our results as a function of the rapidity $y$ at $\sqrt{s_{NN}}=5.02$ TeV and the invariant 
collision energy $\sqrt{s_{NN}}$ at $y=0$.   
 We also give predictions
for the incoherent photoproduction cross section $\gamma Pb \to \rho A^{\prime}$ as a 
function of the invariant photon--nucleon energy $W_{\gamma p}$. 
 We demonstrate that the effect of the inelastic nuclear shadowing in the 
incoherent cross sections is significant and leads to an additional 25\% suppression of the cross section.
Comparing our predictions to those of the STARlight Monte Carlo framework, we find very significant differences.

\end{document}